\documentclass{revtex4}

\usepackage{amsmath}
\usepackage{graphicx}
\usepackage{amsfonts}
\usepackage{amssymb}
\usepackage{amsthm}
\usepackage{mathrsfs}

\usepackage{hyperref}
\hypersetup{colorlinks=true, linkcolor=blue, citecolor=blue, filecolor=blue, urlcolor=blue,
            pdfproducer={Latex},
            pdfcreator={pdflatex}}
\newcommand{\fracc}{\displaystyle\frac}

\begin{document}

\title {Fractional derivative models for atmospheric dispersion of pollutants}

\author{A. G. O. Goulart$^1$, M.~J. Lazo$^1$, J. M. S. Suarez$^1$, D. M. Moreira$^2$}

\affiliation{$^1$Instituto de Matem\'atica, Estat\'{\i}stica e F\'{\i}sica, Universidade Federal do Rio Grande, 96.201-900 Rio Grande, Rio Grande do Sul, Brazil}

\affiliation{$^2$ Center for Integrated Manufacturing and Technology, SENAI/CIMATEC, Brazil}

\begin{abstract}

In the present work, we investigate the potential of fractional derivatives to model atmospheric dispersion of pollutants. We propose simple fractional differential equation models for the steady state spatial distribution of concentration of a non-reactive pollutant in Planetary Boundary Layer. We solve these models and we compare the solutions with a real experiment. We found that the fractional derivative models perform far better than the traditional Gaussian model and even better than models found in the literature where it is considered that the diffusion coefficient is a function of the position in order to deal with the anomalous diffusion.
\end{abstract}

\maketitle

\section{Introduction}

The dispersion of pollutants in the atmosphere is a permanent source of challenging problems due to its physical complexity. An example of nontrivial problems is the description of diffusion under atmospheric turbulence. Indeed, the turbulence is the reason behind the dispersion of pollutants in the atmosphere, since without turbulence the pollutants would follow only the streamlines of mean wind velocities, displaying minimal diffusion in other directions.
A remarkable consequence of turbulence is the emergence of anomalous diffusion. Unlike common diffusion, where the mean square displacement increases linearly with time, in the anomalous diffusion the mean square displacement is not linear. The anomalous diffusion is closely connected with the failure of the central limit theorem due to sparse distribution or long-range correlations. Actually, the anomalous diffusion is related to the more general L\'{e}vy--Gnedenko theorem that generalizes the central limit theorem for situations where not all moments exist \cite{Metzler}.
Historically, anomalous diffusion was first observed in nature in the pollutant dispersion phenomenon. In 1926, Richardson measured the increase in the width of plumes of smoke generated from point sources located in a  turbulent velocity field \cite{Metzler,Metzler2,Richardson,West}. Based on its observations, Richardson speculated that the speed of turbulent air, which has a non-differentiable structure, can be approximately described by the Weierstrass function. This was motivated partly by the observation that the width of smoke plumes grows with $t^{\alpha}$ ($\alpha\geq 3$), unlike common diffusion where $\alpha=1$. Furthermore, the non-differentiable behavior of the width growth of plumes generated from a point source is directly related to the fractal structure of the turbulent velocity field, where the fluctuation's size scales are, in many cases, very large compared to the average scale.

In this context, traditional differential equations do not adequately describe the problem of turbulent diffusion because usual derivatives are not well defined in the non-differentiable behavior introduced by turbulence. Consequently, it is expected that classical advection-diffusion equations do not fully explain the anomalous diffusion of pollutants since in this case the parameters of the system usually grow faster than the solutions obtained by classical models \cite{West}. However, despite this facts, traditional Eulerian and Lagrangian models are the most frequently used in modeling the dispersion of contaminants in the atmosphere \cite{GL,HP,MRVG,Wilson}. The Eulerian models basically consist of the solution of the advection-diffusion equation and the Lagrangian models are based on the solution of the Langevin equation. In order to deal with the anomalous diffusion, a common procedure used in the literature in order to modify Eulerian models is to assume that the physical structure of turbulent flow and velocity fields are described by a complex diffusion coefficient and mean velocity profile that are both considered as functions of spatial coordinates. These functions are usually chosen in order to fit experimental data or are obtained from Taylor statistical diffusion theory \cite{Taylor,Batchelor,DACMT,GMCT}.

The dispersion model that we propose in the present work, unlike the models usually found in the literature, does not solve the advection-diffusion equation as expressed traditionally, but modifies the mathematical structure of this equation to represent more realistically the evolution in space of concentration of contaminants dispersed in a turbulent flow. In this sense, we introduce fractional operators in the equation that governs the distribution of contaminants in the atmosphere. The use of fractional calculus in the modeling of turbulent diffusion is justified by the non-differentiable behavior in the problem and by the presence of anomalous diffusion. In the last decades, thousands of works were carried out with the objective of explaining anomalous diffusion. However, few works deal with the analysis of the validity of models based on classical differential equations, and or the use of unusual differential operators, to describe systems displaying non-differential behavior and/or anomalous dynamics. In this subject, we can highlight the use of fractional calculus that emerged as an valuable mathematical tool to model the time evolution of anomalous diffusion since fractional derivatives arouses naturally in anomalous diffusion process (a detailed physical explanation for the emergence of fractional derivatives in anomalous diffusion models can be found in the reviews \cite{Metzler,Metzler2,West}). However, the use of fractional derivatives in the study of steady states systems, and in particular the dispersion of pollutants in the atmosphere, is underexplored. In this context, in order to investigate the potential of application of fractional derivatives to modeling dispersion of pollutants, we propose two very simple fractional differential equation models for the spatial distribution of concentration of a non-reactive pollutant in Planetary Boundary Layer (PBL). We solve these models and we compare the solutions with traditional integer order derivative models and with a real experiment.

The paper is organized in the following way. In Section II we review the basic definitions and properties of Riemann-Liouville and Caputo fractional derivatives, that are needed for formulating our fractional differential equation model. The models are presented and analytically solved in Section III. A numerical comparison with our solution against traditional models and experimental data is done in Section IV. Finally, Section V presents the conclusions.


\section{Mathematical background}

There are several definitions of fractional order derivatives. Theses definitions include the Riemann-Liouville, Caputo, Riesz, Weyl, Grunwald-Letnikov, etc. (see \cite{OldhamSpanier,Kilbas,SKM,MillerRoss,Podlubny,Diethelm} for a review of fractional calculus and fractional differential equations). In this section, we review the basic definitions and properties of the Riemann-Liouville and  Caputo Fractional Calculus. Despite we have many different approaches to fractional calculus, several known formulations are somehow connected with the analytical continuation of Cauchy formula for $n$-fold integration,
that give us a definition for an integration $\!\!\!\!\!\phantom{J}_a J^{\alpha}_x$
of noninteger (or fractional) order defined by:
\begin{equation}
\label{a3}
\!\!\!\!\!\phantom{J}_a J^{\alpha}_x f(x) =\frac{1}{\Gamma(\alpha)}\int_{a}^x \frac{f(u)}{(x-u)^{1-\alpha}}du \;\;\;\;\; (\alpha \in \mathbb{R}_+)
\end{equation}
with $a<b$ and $a,b\in \mathbb{R}$.
This integral is historically called left fractional Riemann-Liouville integral of order $\alpha \in \mathbb{R}$. For integer $\alpha$ the fractional Riemann-Liouville integral \eqref{a3} coincides with the usual integer order $n$-fold integration. Moreover, from the definition \eqref{a3} it is easy to see that the Riemann-Liouville fractional integral converges for any integrable function $f$ if $\alpha>1$. Furthermore, it is possible to proof the convergence of \eqref{a3} for $f\in L_1[a,b]$ even when $0<\alpha<1$ \cite{Diethelm}.

The fractional order integration \eqref{a3} is the building block of the Riemann-Liouville and Caputo calculus, the two most popular formulations of fractional calculus, as well as several other approaches to fractional calculus. The left Riemann-Liouville fractional derivative of order $\alpha >0$ ($\alpha\in \mathbb{R}$) is defined by ${_a D^{\alpha}_x} f(x) = D^{n}_x \!\!\!\!\phantom{J}_a J^{n-\alpha}_x  f(x)$ with $n=[\alpha]+1$, namely:
\begin{equation}
\label{a7}
{_a D^{\alpha}_x} f(x)=\frac{1}{\Gamma(n-\alpha)}\frac{d^n}{dx^n}\int_{a}^x \frac{f(u)}{(x-u)^{1+\alpha-n}}du \;\;\;\;\; (n=[\alpha]+1,\; \alpha\in \mathbb{R}_+^*, \; a\in \mathbb{R}).
\end{equation}
On the other hand, the left Caputo fractional derivative of order $\alpha >0$ ($\alpha\in \mathbb{R}$) is defined by ${_a^C D^{\alpha}_x} f(x) = \!\!\!\!\phantom{J}_a J^{n-\alpha}_x D^{n}_x f(x)$ with $n=[\alpha]+1$, namely:
\begin{equation}
\label{a8}
{_a^C D^{\alpha}_x} f(x)=\frac{1}{\Gamma(n-\alpha)}\int_{a}^x \frac{f^{(n)}(u)}{(x-u)^{1+\alpha-n}}du \;\;\;\;\; (n=[\alpha]+1,\; \alpha\in \mathbb{R}_+^*, \; a\in \mathbb{R}),
\end{equation}
where $f^{(n)}(u)=\frac{d^n f(u)}{du^n}$ are ordinary derivatives of integer order $n$, and $f^{(n)}\in L_1[a,b]$.

An important consequence of definitions \eqref{a7} and \eqref{a8} is that the Riemann-Liouville and Caputo fractional derivatives are nonlocal operators. The left differ-integration operators \eqref{a7} and \eqref{a8} depend on the values of the function at left of $x$, i.e. $a\leq u \leq x$ ($x\leq u \leq b$). Moreover, when $\alpha$ is an integer the Riemann-Liouville and Caputo derivatives reduce to ordinary derivatives of order $\alpha$.
It is important to notice the remarkable fact that while the Riemann-Liouville derivative of a constant is not zero, the Caputo derivatives of a constant is zero for all $\alpha>0$.
Furthermore, for power functions $(x-a)^{\beta}$ and $(b-x)^{\beta}$ ($\beta>0$) we have:
\begin{eqnarray}
\label{a10b}
{_a D^{\alpha}_x} (x-a)^{\beta}=\frac{\Gamma(\beta+1)}{\Gamma(\beta+1-\alpha)}(x-a)^{\beta-\alpha},\\ \nonumber
{_a^C D^{\alpha}_x} (x-a)^{\beta}=\frac{\Gamma(\beta+1)}{\Gamma(\beta+1-\alpha)}(x-a)^{\beta-\alpha}.
\end{eqnarray}

Finally, let us consider now a real valued function $f:\Omega \rightarrow \mathbb{R}$ of $d$ real variables $x_1,x_1,...x_d$ defined over the domain $\Omega = [a_1,b_1]\times \cdots \times [a_d,b_d]\subset \mathbb{R}^{d}$. We can define the left Riemann-Liouville and Caputo partial fractional derivatives of order $\alpha\in \mathbb{R}_+^*$ with respect to $x_i$  ($n=[\alpha]+1$, $a_i,b_i\in \mathbb{R}$) as
\begin{equation}
\label{aa1}
\!\!\!\!\! \frac{\partial^{\alpha}}{\partial x_i^\alpha} f(x_1,...,x_d)=\frac{1}{\Gamma(n-\alpha)}\frac{\partial^{n}}{\partial x_i^n}\int_{a_i}^{x_i} \frac{f(x_1,...,x_{i-1},u,x_{i+1},...,x_d)}{(x_i-u)^{1+\alpha-n}}du
\end{equation}
and
\begin{equation}
\label{aa8}
\!\!\!\!\! \frac{^C\partial^{\alpha}}{\partial x_i^\alpha} f(x_1,...,x_d)=\frac{1}{\Gamma(n-\alpha)}\int_{a_i}^{x_i} \frac{\partial_{u}^{n}f(x_1,...,x_{i-1},u,x_{i+1},...,x_d)}{(x_i-u)^{1+\alpha-n}}du
\end{equation}
respectively, where $\partial_u^n$ is the ordinary partial derivative of integer order $n$ with respect to the variable $u$.


\section{The mathematical models}

An equation for the spatial distribution of concentration of a non-reactive pollutant in PBL can be obtained by an application of the principle of continuity or conservation of mass, where the flows are represented by the K-theory \cite{Csanady,Blackadar}. In a Cartesian coordinate system in which the longitudinal direction $x$ coincides with the average wind velocity, the spatial distribution of concentration  $\bar{c}=\bar{c}(x,y,z)$ of a non-reactive substance can be described by the advection-diffusion equation
\begin{equation}\label{eq 1}
 u\frac{{\partial}\;\bar{c}}{\partial x}=\frac{\partial}{\partial x}(K_{x}\frac{\partial \; \bar{c}}{\partial x})+\frac{\partial}{\partial y}(K_{y}\frac{\partial \; \bar{c}}{\partial y})+\frac{\partial}{\partial z}(K_{z}\frac{\partial \; \bar{c}}{\partial z}),
\end{equation}
where we consider only the steady state case,  $u=u(z)$ is the mean wind speed in the longitudinal direction and $K_{x}$, $K_{y}$, $K_{z}$ are the diffusion coefficient. The equation of the cross-wind integrated concentration ($\overline{c^{y}}=\overline{c^{y}}(x,z)$) is obtained by integrating the Eq. (\ref{eq 1}) with respect to $y$ from $- \infty$ to $+ \infty$ (neglecting the longitudinal diffusion),
\begin{equation}\label{eq 2}
 u\frac{{\partial}\;\overline{c^{y}}}{\partial x}=\frac{\partial}{\partial z}(K_{z}\frac{\partial \; \overline{c^{y}}}{\partial z}).
\end{equation}
A process governed by the steady state regime of the advection-diffusion equation \eqref{eq 2} with $K_z$ constant is called a Gaussian process with a sharp boundary condition $\overline{c^{y}}_0(z)=\lim_{x\rightarrow 0^+}\overline{c^{y}}(x,z)=\delta(z)$ in the domain $0\leq x< \infty$ and $-\infty<z<\infty$ the solution of \eqref{eq 2} is given by the Gaussian shape
\begin{equation}
\label{eq 2b}
\overline{c^{y}}(x,z)=\frac{1}{\sqrt{4\pi K_z x}}\exp \left(\frac{-z^2}{4K_z x}\right).
\end{equation}
On the other hand, by taking a Fourier transform of \eqref{eq 2} we obtain a relaxation equation for a fixed wavenumber $k$,
\begin{equation}
\label{eq 2c}
u\frac{{\partial}\;\overline{c^{y}}(x,k)}{\partial x}=-K_{z} k^2 \overline{c^{y}}(x,k),
\end{equation}
which implies that individual modes of \eqref{eq 2} on rectangular domains decay exponentially in $x$
\begin{equation}
\label{eq 2d}
\overline{c^{y}}(x,k)=\overline{c^{y}}(0,k)\exp \left(-K_{z} k^2 x \right),
\end{equation}
where we set for simplicity $u=1$. Consequently, for a Gaussian process, individual modes of the concentration follows an exponential distribution in the longitudinal direction $x$. Actually, the Gaussian distribution for the concentration \eqref{eq 2b} and its individual modes exponential distribution \eqref{eq 2d} are characteristic of the normal diffusive process, displaying a linear mean squared displacement
\begin{equation}
\label{eq 2e}
\langle z^2(x) \rangle =-\lim_{k\rightarrow 0}\frac{d^2}{dk^2}\overline{c^{y}}(x,k)=2K_z x.
\end{equation}
On the other hand, in most cases, the anomalous process displays a power law mean squared displacement $\langle z^2 \rangle \propto x^{\alpha}$ with $0< \alpha <1$ \cite{Metzler}. This power law follows directly from the Fourier-Laplace concentration function \cite{Metzler}
\begin{equation}
\label{eq 2f}
\overline{c^{y}}(s,k)=\frac{\overline{c^{y}}(0,k)}{s+K_z s^{1-\alpha}k^2},
\end{equation}
since
\begin{equation}
\label{eq 2g}
\langle z^2(x) \rangle =-\mathscr{L}^{-1} \left\{ \lim_{k\rightarrow 0}\frac{d^2}{dk^2}\overline{c^{y}}(s,k)\right\}=\frac{2K_z}{\Gamma(\alpha+1)} x^{\alpha},
\end{equation}
where $\Gamma(\cdot)$ is the Gamma function and $\mathscr{L}^{-1}$ stands for the inverse Laplace transform. Moreover, from the Laplace transform for fractional Riemann-Liouville integrals \cite{OldhamSpanier,SKM}
\begin{equation}
\label{eq 2h}
\mathscr{L} \left\{ {_0 J^{\alpha}_x} f(x)\right\}=s^{-\alpha}f(s),
\end{equation}
we can infer from \eqref{eq 2f} the steady-state fractional advection-diffusion integral equation
\begin{equation}
\label{eq 2i}
\overline{c^{y}}(x,z)-\overline{c^{y}}(0,z)={_0 J^{\alpha}_x} K_z \frac{\partial^ 2 \overline{c^{y}}(x,z)}{\partial z^ 2},
\end{equation}
for $u=1$. Here and in the rest of the work we set without loss of generality $a=0$. Furthermore, by taking the derivative of \eqref{eq 2i} in $x$ we obtain a Riemann-Liouville fractional differential equation
\begin{equation}
\label{eq 2j}
\frac{\partial \overline{c^{y}}(x,z)}{\partial x}=\frac{\partial^{1-\alpha}}{\partial x^{1-\alpha}}  K_z \frac{\partial^ 2 \overline{c^{y}}(x,z)}{\partial z^ 2}.
\end{equation}
Although it is well established that fractional differential equations describe more realistically the time evolution of anomalous diffusive systems \cite{Metzler,Metzler2,West}, in the present work we shall show that the fractional differential equation \eqref{eq 2j} is more adequate than the classical equation \eqref{eq 2} to describes the steady state regime of anomalous dispersion of pollutants in a turbulent flow. Before proceeding the work, let us rewrite \eqref{eq 2j} in a more adequate expression. Given that for continuum functions $f(x)$ we have ${_0 J^{\alpha}_x}{_0D_x^\alpha}f(x)={_0D_x^\alpha}{_0 J^{\alpha}_x}f(x)=f(x)$, and ${_0 J^{\alpha}_x}{_0 J^{\beta}_x}={_0 J^{\alpha+\beta}_x}$ \cite{OldhamSpanier,Kilbas,SKM,Diethelm}, we obtain after some manipulations
\begin{equation}
\label{eq 2k}
\frac{\partial^{\alpha}\overline{c^{y}}(x,z)}{\partial x^{\alpha}} -\frac{x^{-\alpha}}{\Gamma(1-\alpha)}\overline{c^{y}}(0,z)= K_z \frac{\partial^ 2 \overline{c^{y}}(x,z)}{\partial z^ 2}.
\end{equation}
Finally, the Riemann-Liouville derivative in \eqref{eq 2k} can be replaced by a Caputo derivative since they are directly related
$$
{_0^CD_x^\alpha}f(x)={_0D_x^\alpha}f(x)- \frac{f(0)}{\Gamma(1-\alpha)}x^{-\alpha}
$$
for $0<\alpha<1$ \cite{Kilbas,Diethelm}. Consequently, the general steady-state fractional advection-diffusion equation displaying the anomalous power law mean squared displacement \eqref{eq 2g} is given by
\begin{equation}
\label{eq 2l}
u\frac{{^C\partial^{\alpha}}\overline{c^{y}}(x,z)}{\partial x^{\alpha}} = K_z \frac{\partial^ 2 \overline{c^{y}}(x,z)}{\partial z^ 2},
\end{equation}
for $u$ constant. The practical advantage of \eqref{eq 2l} instead of the equivalent equation \eqref{eq 2j} is that while differential equations containing Caputo derivatives requires regular physical boundary conditions, differential equations containing Riemann-Liouville derivatives requires non-regular boundary conditions that are more difficult to implement and to physically interpret \cite{HeymansPodlubny}.

Actually, in traditional atmospheric dispersion models, the parameters $u$ and $K_z$ are usually considered as functions on $x$ and $z$ in order to deal with anomalous diffusion introduced by turbulence. In this context, the turbulent parameterization is related to the estimate of the mean wind speed and the diffusion coefficient \cite{Moreira,Sharan}. The wind speed is obtained from the similarity theory \cite{Panofsky,Smith,Irwin} and the diffusion coefficient is obtained from Taylor statistical diffusion theory \cite{Taylor,Batchelor}.
In this work is used the power law profile for mean wind speed \cite{Smith,Irwin},
\begin{equation}\label{eq 3}
u(z)=\gamma z^{k},\;\;\; \gamma=\frac{u(z_{1})}{z_{1}^{k}}
\end{equation}
where $u(z_{1})$ is the wind speed at known height $z_{1}$ and $k$ is the power law exponent dependent on the intensity of turbulence.
In order to obtain a simple analytic solution of equation \eqref{eq 2k}, the diffusion coefficient can be specified as a function of downwind distance. We consider here two important cases. The first when $u=\gamma$ and $K_{z}=K$ are constants give us a fractional generalization of the Gaussian model (that we shall call $\alpha$--Gaussian model)
\begin{equation}\label{eq 16}
\gamma\frac{{^C{\partial}^{\alpha}}\overline{c^{y}}}{\partial x^{\alpha}}=K \frac{\partial^{2} \; \overline{c^{y}}}{\partial z^{2}}.
\end{equation}
The second case is a fractional generalization of the model proposed by Sharan and Modani (SM model) \cite{Sharan}, where now we have
\begin{equation}\label{eq 4}
K_{z}=K(x)=\rho  x^{\alpha}, \;\;\; \rho=\nu u,
\end{equation}
where $\nu = (\frac{\sigma_{w}}{u})^{2}$ is the turbulence parameter \cite{Arya}, $\sigma_{w}$ is the standard deviation of the vertical component of the wind speed,  and $0< \alpha<1$ is the same as the order of the fractional derivative in \eqref{eq 2}.
Substituting \eqref{eq 3} and \eqref{eq 4} into \eqref{eq 2k} we define the $\alpha$--SM model
\begin{equation}\label{eq 5}
\gamma z^{k}\frac{{^C{\partial}^{\alpha}}\overline{c^{y}}}{\partial x^{\alpha}}=\rho x^{\alpha} \frac{\partial^{2}\overline{c^{y}}}{\partial z^{2}}.
\end{equation}
It is important to remark that the diffusion coefficient $K_{z}=K(x)=\rho  x$ considered in the SM model \cite{Sharan} (and \eqref{eq 4} in $\alpha$--SM model) is not physically meaningful in the context of diffusion of pollutants in real atmospheric condition. This diffusion coefficient is introduced in \cite{Sharan} only in order to fit experimental data with a simple classical advection-diffusion equation. In the present work, we analyzed the $\alpha$--SM model \eqref{eq 5} with \eqref{eq 4} only in order to investigate if there is any advantage, against the $\alpha$--Gaussian model \eqref{eq 16}, of considering the diffusion coefficient as a function of spatial coordinates.

Finally, in order to \eqref{eq 2k}, \eqref{eq 16} and \eqref{eq 5} describe a possible real problem of dispersion in PBL, it should be imposed boundary conditions of zero flux on the ground ($z=z_{0}$) and top ($z=h)$, and the pollutant is released from an elevated point source of emission rate $Q$  at height $H_{s}$, i.e.,
\begin{equation}\label{eq 6}
K_{z}\frac{\partial \overline{c^{y}} }{\partial z}=0,\;\;\;\;z=z_{0},\;\;z=h,
\end{equation}
\begin{equation}\label{eq 7}
u \overline{c^{y}}(0,z)=Q\delta(z-H_{s}),\;\;x=0,
\end{equation}
where $z_{0}$ is the surface roughness length, $h$ is the PBL height and $\delta(\cdot)$ is the Dirac delta function.

Equations \eqref{eq 16} and \eqref{eq 5} can be analytically solved by using the separation of variables technique. We begin by assuming that both \eqref{eq 16} and\eqref{eq 5} have a solution of the form
\begin{equation}\label{eq 8}
\overline{c^{y}}(x,z)=X(x)Z(z).
\end{equation}
Substituting \eqref{eq 8} in \eqref{eq 16}, we get two ordinary differential equations in the variables $X$ and $Z$ as follows,
\begin{equation}\label{eq 9a}
{_0^C D^{\alpha}_x} X+\kappa \lambda^{2} X=0,
\end{equation}
and
\begin{equation}\label{eq 10a}
\frac{d^{2}Z}{d z^{2}}+\lambda^{2} Z=0.
\end{equation}
where $\kappa = \frac{K}{\gamma}$. On the other hand, by substituting \eqref{eq 8} in \eqref{eq 5} we get
\begin{equation}\label{eq 9}
{_0^C D^{\alpha}_x} X+\tau \lambda^{2} x^{\alpha} X=0,
\end{equation}
and
\begin{equation}\label{eq 10}
\frac{d^{2}Z}{d z^{2}}+\lambda^{2} z^{k}Z=0.
\end{equation}
where $\tau = \frac{\rho}{\gamma}$.

\subsection{The solution of the $\alpha$--Gaussian model \eqref{eq 16}}
The solution of \eqref{eq 16} subject to the boundary conditions \eqref{eq 6} and \eqref{eq 7} can be obtained from \eqref{eq 9a} by the Frobenius method. The solution for the $\alpha$--Gaussian model \eqref{eq 16} with $z_0 \approx 0$ is (see Appendix A)
\begin{equation}\label{eq 19}
\overline{c^{y}}(x,z)=\frac{Q}{u h}\Big[1+ 2 \sum_{n=1}^{\infty}\cos( \lambda_{n} H_{s})\cos(\lambda_{n} z)E_{\alpha}\Big(-\kappa \lambda_{n}^{2}x^{\alpha}\Big)\Big],
\end{equation}
where $\lambda_n=\frac{n\pi}{h}$, and
\begin{equation}\label{eq 18}
E_{\alpha}(x)=\sum_{n=0}^{\infty}\frac{x^{n}}{\Gamma(n \alpha + 1)}
\end{equation}
is the Mittag-Leffler function.
The solution for the classical Gaussian model is obtained from equation \eqref{eq 16} when $ \alpha = 1$ \cite{Csanady}
\begin{equation}\label{eq 20}
\overline{c^{y}}(x,z)=\frac{Q}{u h}\Big[1+ 2 \sum_{n=1}^{\infty}\cos( \lambda_{n} H_{s})\cos(\lambda_{n} z)\exp \Big(-\kappa \lambda_{n}^{2}x\Big)\Big].
\end{equation}

The big advantage of fractional differential equations, against integer order ones, to describe anomalous diffusion can easily be seen from the solution of the $\alpha$-Gaussian model \eqref{eq 19}. Even for a constant diffusion coefficient, the concentration of pollutants in \eqref{eq 19} falls more slowly in $x$ compared with the traditional Gaussian model \eqref{eq 20}, since the Mittag-Leffler function in \eqref{eq 19} for $0< \alpha < 1$ falls more slowly than the exponential function in \eqref{eq 20}. Actually, this was already expected since the $\alpha$--Gaussian model displays a power law mean squared displacement \eqref{eq 2g}. Furthermore, the function $E_{\alpha}(-\kappa \lambda_{n}^{2}x^{\alpha})$ appearing in the solution of the $\alpha$-Gaussian model \eqref{eq 19} is just related to the Mittag-Leffler distribution $1-E_{\alpha}(-\kappa \lambda_{n}^{2}x^{\alpha})$ that is characteristic of anomalous diffusion process like the fractional Brownian motion \cite{Metzler,Metzler2} (it is also well known that the L\'evy distribution, found in L\'evy flight anomalous diffusion, and Mittag-Leffler distribution are related \cite{Feller,HMS}).

\subsection{The solution of the $\alpha$--SM model \eqref{eq 5}}

The solution of the fractional $\alpha$--SM model \eqref{eq 5} subject to the boundary conditions \eqref{eq 6} and \eqref{eq 7} can be obtained from \eqref{eq 9} and \eqref{eq 10} by the Frobenius method. The solution of \eqref{eq 9} is (see Appendix B)
\begin{equation}\label{eq 11}
X(x)=\sum_{j=0}^{\infty} (-1)^{j}\frac{\prod_{i=1}^{j}\Gamma[(2i-1)\alpha+1]}{\prod_{i=1}^{j+1}\Gamma[2(i-1)\alpha+1]}(\tau \lambda^{2}x^{2\alpha})^{j}
\end{equation}
On the other hand, the solution of \eqref{eq 10} is (Appendix B),
\begin{equation}\label{eq 12}
Z(z)=z^{\frac{1}{2}}\Big[A J_{\mu}(2 \lambda \mu z^{\frac{1}{2\mu}})+B J_{-\mu}(2 \lambda \mu z^{\frac{1}{2\mu}})\Big]
\end{equation}
where $J_{\mu}$ and $J_{-\mu}$ are the Bessel functions of the first kind and order $\mu$ and $-\mu$, respectively, and $\mu=\frac{1}{2+k}$. Applying the boundary condition given by \eqref{eq 6} at $z=z_{0}$ (with $z_{0}\approx 0$), we get (Appendix B)
\begin{equation}\label{eq 13}
A=0.
\end{equation}
From the boundary condition given by \eqref{eq 6} at $z=h$, we obtain the following relationship (Appendix B),
\begin{equation}\label{eq 14}
J_{-\mu+1}(2\lambda \mu h^{\frac{1}{2\mu}})=0.
\end{equation}
Therefore, the $\lambda = \lambda_{n}$ are the zeros of the Bessel function of order $-\mu + 1$.
Substituting \eqref{eq 11}, \eqref{eq 12}, \eqref{eq 13} and \eqref{eq 14} in \eqref{eq 8} we obtain the equation for the cross-wind integrated concentration of a non-reactive pollutant in PBL released from an elevated point source
\begin{equation}\label{eq 15}
\overline{c^{y}}(x,z)=B_{0}+z^{\frac{1}{2}}\sum_{n=1}^{\infty}B_{n}J_{-\mu}(2 \lambda_{n} \mu z^{\frac{1}{2\mu}})\sum_{j=1}^{\infty} (-1)^{j}\frac{\prod_{i=1}^{j}\Gamma[(2i-1)\alpha+1]}{\prod_{i=1}^{j+1}\Gamma[2(i-1)\alpha+1]}(\tau \lambda_{n}^{2}x^{2\alpha})^{j}.
\end{equation}
The constants $B_{0}$ and $B_{n}$ are obtained from the boundary condition given by \eqref{eq 7}, and $\lambda_{n} $ are given by the zeros of the Bessel function of order $-\mu + 1$ calculated from \eqref{eq 14}. The function \eqref{eq 15} is the solution of our generalized model \eqref{eq 5} obtained when $u$ and  $K_{z}$ are functions of the $z$ ( \eqref{eq 3} and \eqref{eq 4}).


\section{Performance of the models}

The performance of the fractional derivative models expressed in \eqref{eq 19} and \eqref{eq 15} was evaluated against experimental ground-level concentration using the Copenhagen dispersion experiment \cite{Briggs1973}. The experiment took place in Copenhagen in $1978$. The tracer was released without buoyancy at a height of $115m$ and collected at ground level positions at a maximum of three crosswind arcs of tracer sampling units. The site was mainly residential with a roughness length of the $0.6m$ \cite{Gryning,Gry-Lyck}. The meteorological conditions of the Copenhagen dispersion experiment are shown in Table $1$, where $\bar{u}_{10}$ is the mean wind speed at $10m$, $u_{*}$ is the friction velocity and $L$ is the Monin–-Obukhov length.

\begin{table}[!h]
\centering
\caption{Meteorological conditions during the Copenhagen experiment}
\begin{tabular}{ lccccccc}
\hline
Exp.   & $\bar{u}_{10} (ms^{-1})$ & $u_{*} (ms^{-1})$ & L(m)  \;\;\;\; & $\sigma_{w}(ms^{-1} )$  & $h$ (m)   \\
\hline
 1     & 2.1                 & 0.37              & -46    \;\;\;\;& 0.83                   & 1980     \\
 2     & 4.9                 & 0.74              & -384   \;\;\;\;& 1.07                   & 1920     \\
 3     & 2.4                 & 0.39              & -108   \;\;\;\;& 0.68                   & 1120     \\
 4     & 2.5                 & 0.39              & -173   \;\;\;\;& 0.47                   & 390     \\
 5     & 3.1                 & 0.46              & -577   \;\;\;\;& 0.71                   & 820     \\
 6     & 7.2                 & 1.07              & -569   \;\;\;\;& 1.33                   & 1300     \\
 7     & 4.1                 & 0.65              & -136   \;\;\;\;& 0.87                   & 1850     \\
 8     & 4.2                 & 0.70              & -72    \;\;\;\;& 0.72                   & 810     \\
 9     & 5.1                 & 0.77              & -382   \;\;\;\;& 0.98                   & 2090     \\
\hline
\end{tabular}
\end{table}

Usually, the performance of dispersion models is evaluated from a well know set of statistical indices described by Hanna \cite{Hanna} defined in the following way,
\begin{equation*}
\begin{split}
\text{NMSE}\; (\text{normalized mean square error})&= \frac{\overline{(c_{o}-c_{p})^{2}}}{\overline{c_{o}}\overline{c_{p}}},\\
\text{Cor}\; (\text{correlation coefficient})&= \frac{\overline{(c_{o}-\overline{c_{p}})(c_{p}-\overline{c_{p}})}}{\sigma_{o}\sigma_{p}},\\
\text{FB}\; (\text{fractional bias})&=\frac{\overline{c_{o}-\overline{c_{p}}}}{0.5(\overline{c_{o}}+\overline{c_{p}})},\\
\text{FS}\; (\text{fractional standard deviations}) &= \frac{\sigma_{o}-\sigma_{p}}{0.5(\sigma_{o}+\sigma_{p})},
\end{split}
\end{equation*}
where $c_{p}$ is the computed concentration, $c_{o}$ is the observed concentration, $\sigma_{p}$ is the computed standard deviation, $\sigma_{o}$ is the observed standard deviation, and the overbar indicates an averaged value. The statistical index FA2 represents the fraction of data for $0.5\leq \frac{c_{p}}{c_{o}} \leq 2 $. The best results are indicated by values nearest to $0$ in NMSE, FS and FB, and nearest to $1$ in Cor and FA2. The statistical indices used to evaluate our model's performance is shown in Tables $3$.

\begin{table}[!h]
\centering
\caption{\small Observed and estimated crosswind-integrated concentrations $\frac{\overline{c^{y}}}{Q}$($10^{-4}s m^{-2}$) for Copenhagen experiment.}
\begin{tabular}{lccccccc}
\hline
   Exp.                   & Distance (m)       & Observed       &  $\alpha$--Gaussian    & Gaussian    & $\alpha$--SM & SM     \\
\hline
 1                        &     1900           &   6.48         &     6.32               &    3.61              &     7.74              &   4.62            \\
 1                        &     3700           &   2.31         &     4.97               &    2.72              &     4.20              &   2.30            \\
 2                        &     2100           &   5.38         &     4.14               &    2.47              &     4.89              &   3.18            \\
 2                        &     4200           &   2.95         &     3.27               &    1.76				 &     2.79              &   1.58            \\
 3                        &     1900           &   8.20         &     6.51               &    4.00              &     8.91             &   5.66            \\
 3                        &     3700           &   6.22         &     5.22               &    3.73              &     5.03             &   2.88            \\
 3                       &     5400           &   4.30         &     4.66               &    3.72              &     3.95              &   2.21            \\
 4                       &     4000           &  11.7          &    10.60               &   10.25              &     8.21              &   5.93            \\
 5                       &     2100           &   6.72         &     5.71               &    3.98              &     7.55              &   4.81            \\
 5                       &     4200           &   5.84         &     4.70               &    3.93              &     4.28              &   2.43            \\
 5                       &     6100           &   4.97         &     4.36               &    3.93              &     3.27              &   2.00            \\
 6                       &     2000           &   3.96         &     2.90               &    1.72              &     3.90              &   2.63            \\
 6                       &     4200           &   2.22         &     2.27               &    1.24              &     2.23              &   1.28            \\
 6                       &     5900           &   1.83         &     2.08               &    1.12              &     1.79              &   0.90            \\
 7                       &     2000           &   6.70         &     4.68               &    2.77              &     6.21              &   4.16            \\
 7                       &     4100           &   3.25         &     3.65               &    1.95              &     3.50              &   2.03            \\
 7                       &     5300           &   2.23         &     3.34               &    1.73              &     2.79              &   1.56            \\
 8                       &     1900           &   4.16         &     5.75               &    3.51              &     7.32              &   4.87            \\
 8                       &     3600           &   2.02         &     4.72               &    3.01              &     4.68              &   2.74            \\
 8                       &     5300           &   1.52         &     4.18               &    2.95              &     3.39              &   1.84            \\
 9                       &     2100           &   4.58         &     3.77               &    2.26              &     5.26              &   3.44            \\
 9                       &     4200           &   3.11         &     2.99               &    1.61              &     3.07              &   1.74            \\
 9                       &     6000           &   2.59         &     2.63               &    1.35              &     2.24              &   1.19            \\
  \hline
\end{tabular}

\end{table}

\begin{table}[!h]
\centering
\caption{\small Statistical indices to evaluate the performance of proposed models (Eq. \ref{eq 15})}
\begin{tabular}{lccccccc}
\hline
Model                                 & \; Cor \;   & \; NMSE \;    & \; FS \;    & \; FB \;     & \; FA2 \; \\
\hline
  $\alpha$--Gaussian    & 0.83   & 0.07     & 0.30   & 0.001   & 0.87     \\
 Gaussian     & 0.82   & 0.23     & 0.27  & -0.39   & 0.73 \\
 $\alpha$--SM  & 0.83   & 0.08     & 0.17   & 0.03   & 0.83     \\
 SM                     & 0.80   & 0.30     & 0.50   & -0.44   & 0.74     \\
  \hline
\end{tabular}

\end{table}

In Table $2$, the results of the estimated crosswind-integrated concentrations for the Copenhagen experiment \cite{Briggs1973} obtained for the fractional models are compared with experimental data, with the Gaussian model \eqref{eq 20}, and with results obtained from the SM model \cite{Sharan}  (where it was proposed a model given by \eqref{eq 5} with $\alpha=1$). In Gaussian model \eqref{eq 20} and $\alpha$--Gaussian model \eqref{eq 19} the mean wind speed in longitudinal direction ($u$) is obtained from Table 1 and the diffusion coefficient constant $K$ is chosen, in order of comparison, as being the mean of the diffusion coefficient given by equation \eqref{eq 4} (with $\alpha = 1$) over the longest distance $x$ of each experiment.

The good performance of the proposed fractional models is shown in Tables $2$, $3$. In special, Table $3$ shows the statistical indices obtained from the mathematical models and the results obtained in the Copenhagen dispersion experiment \cite{Briggs1973}. The $\alpha$--Gaussian model \eqref{eq 19} performs far better than the Gaussian model \eqref{eq 20} to describe the data of Copenhagen experiment, as can be seen from the statistical indices.
The advantage of the $\alpha$--Gaussian model is a consequence of the anomalous diffusion present on turbulent diffusion that results in a distribution with a power law mean squared displacement. It is important to remark, that the $\alpha$--Gaussian model also has better performance than the SM model \cite{Sharan}, where it was considered that the diffusion coefficient is a function $K_{z}=K(x)=\rho  x$ in order to deal with the anomalous diffusion of the problem. Furthermore, from Table 3 we see that the $\alpha$--SM model \eqref{eq 15} has a very small extra performance over the $\alpha$--Gaussian model. This result indicates that there may not be much advantage to consider the diffusion coefficient as a not physically meaningful function.
Finally, in order to estimate the better $\alpha$ value for each model, we analyzed the solutions from $\alpha=0.7$ to $\alpha=0.99$ by steps of $0.01$.
The results obtained from the $\alpha$--SM \eqref{eq 15} and shown in Table 3 are for $\alpha = 0.92$. For this value of $\alpha$, the model \eqref{eq 15} has its best statistical indices. For the $\alpha$--Gaussian model \eqref{eq 19}, the best performance occurs when $\alpha = 0.80$. This difference in the best $\alpha$--value is related to different diffusion coefficients employed in both cases. In the Gaussian model \eqref{eq 20} the diffusion coefficient ($K$) is constant, but in $\alpha$--SM model the diffusion coefficient is a function of the distance $x$ \eqref{eq 15}.
This suggests a correlation between the order of the equation and the quality of diffusion coefficient to the model best describe the observed concentration data. The traditional models ($\alpha=1$) originate from the equation of molecular diffusion (Fick's law), which assumes a Gaussian distribution that displays a linear mean square displacement. Asymmetries related to turbulent flow in the atmosphere are described by the diffusion coefficient. Already the fractional model assumes an anomalous probability distribution with a power law mean squared displacement \eqref{eq 2g}. This anomalous distribution has been proved to be more efficient to describe the motion of particles in a turbulent flow \cite{Metzler}.


\section{Conclusion}

The main objective of the present work is to investigate the potential application of fractional derivatives to model diffusion of pollutants in the atmosphere. The use of fractional calculus in modeling diffusion of pollutants is justified by the presence of anomalous diffusion due to turbulence. In the last decades, thousands of works were carried out with the objective of explaining anomalous diffusion. However, in this subject only a few works deal with the analysis of the validity of models based on classical differential equations, and or the use of unusual differential operators, to describe systems displaying non-differential behavior and/or anomalous dynamics. On the other hand, in the last decades, the use of fractional differential operators emerged as a valuable mathematical tool to model the time evolution of anomalous diffusion \cite{Metzler,Metzler2,West}. However, the use of fractional derivatives to study steady states regimes of dispersion of pollutants in the atmosphere is actually underexplored. In this context, in order to investigate the potential of application of fractional derivatives to model dispersion of pollutants, we propose two simple fractional differential equation models for the spatial distribution of concentration of a non-reactive pollutant in PBL. These equations are obtained by considering a system displaying a distribution with a power law mean squared displacement, instead of a linear one as in normal diffusion.
We solve these models and we compare the solutions with traditional integer order derivative models and with a real experiment. We found that the fractional derivative $\alpha$--Gaussian model \eqref{eq 19} performs far better than the Gaussian model and even better than the SM model \cite{Sharan}, where it was considered that the diffusion coefficient is a function of the position in order to deal with the anomalous diffusion. Most important, the comparison between the $\alpha$--Gaussian model \eqref{eq 19} and the fractional derivative $\alpha$--SM model \eqref{eq 15} indicates that, by using fractional derivatives, there can be little advantage to consider the diffusion coefficient as a function of position, since the $\alpha$--Gaussian model with constant diffusion coefficient gives a very good performance.

Finally, the results obtained from the fractional derivative models motivates further investigations of applications of fractional differential equations to model diffusion of pollutants.

\section*{Acknowledgments}

This work was supported in part by CNPq and CAPES, Brazilian funding agencies.

\appendix

\section{Solution of the $\alpha$--Gaussian model}

In order to solve the equation \eqref{eq 9a} by Frobenius method, we consider the following general power series
\begin{equation}\label{Aeq 10}
\begin{array}{lll}
 X(x^{\alpha})=x^{p}\displaystyle \sum_{j=0}^{\infty}b_{j}\frac{x^{j \alpha}}{\Gamma(j \alpha+\beta)},
\end{array}
\end{equation}
where $0<\alpha < 1$, and $p,\beta$ are constants. The Caputo fractional derivative of \eqref{Aeq 10} is then given by,
\begin{equation}\label{Aeq 11}
\begin{array}{lll}
 {_0^C D^{\alpha}_x} X(x^{\alpha})& = &\frac{b_{0}}{\Gamma(\beta)} {_0^C D^{\alpha}_x} x^{p}  + \displaystyle\sum_{j=1}^{\infty}\frac{b_{j}}{\Gamma(j \alpha+\beta)}{_0^C D^{\alpha}_x} x^{j \alpha + p}\\
                             &  = &\frac{b_{0}}{\Gamma(\beta)}{_0^C D^{\alpha}_x} x^{p}  + \displaystyle\sum_{j=1}^{\infty}\frac{b_{j}}{\Gamma(j \alpha+\beta)} \frac{\Gamma(j \alpha +p+1)}{\Gamma(j \alpha +p-\alpha+1)}x^{(j-1) \alpha + p}.
\end{array}
\end{equation}
Without loss of generality, by considering $\beta = p + 1$ we have,
\begin{equation}\label{Aeq 12}
\begin{array}{lll}
 {_0^C D^{\alpha}_x} X(x^{\alpha}) = \frac{b_{0}}{\Gamma(\beta)}{_0^C D^{\alpha}_x} x^{p} + \displaystyle\sum_{j=0}^{\infty}\frac{b_{j+1}}{\Gamma(j \alpha+\beta)} x^{j \alpha + p},
\end{array}
\end{equation}
where,
\begin{equation} {_0^C D^{\alpha}_x} x^{p} = \left\{\begin{array}{cl}
& 0, \;\;\;\;\;\;\;\;\;\;\;\;\;\;\;\;\;\;\;\;\;\;\;\;\; p=0 \\
& \frac{\Gamma (p+1)}{\Gamma (p+1-\alpha)} x^{p-\alpha}, \;\;\;\;\;\;\;\;0 < p \leq 1.
\end{array} \right.
\end{equation}
If $p < 0$, the function $\frac{dx^p}{dx}$, and consequently $\frac{dX(x^\alpha)}{dx}$, is not integrable and the Caputo fractional derivative is not defined. Substituting equation \eqref{Aeq 12} into equation \eqref{eq 9a} we obtain
\begin{equation}\label{aeq 13}
 \frac{b_{0}}{\Gamma(\beta)}{_0^C D^{\alpha}_x} x^{p} + \displaystyle\sum_{j=0}^{\infty}\Big[b_{j+1} + \kappa \lambda^{2} b_{j}\Big] \frac{x^{j \alpha + p}}{\Gamma(j \alpha + \beta)} = 0.
\end{equation}
If $p\neq 0$ we have the trivial solution $X(x^\alpha)=0$ since for $b_0\neq 0$ the function is not integrable (because in this case $\beta=0,-1,-2,...$ and $p=\beta-1<0$ in order to satisfy \eqref{aeq 13}). For $p=0$, we obtain a nontrivial solution
\begin{equation}
\label{aeq 14}
X(x^\alpha)=b_0 E_{\alpha}\left(-\kappa\lambda^2 x^\alpha\right),
\end{equation}
where $E_\alpha$ is the Mittag-Leffler function \eqref{eq 18}.

On the other hand, the solutions $Z_n(z)$ ($n=0,1,2,3,...$) of \eqref{eq 10a} that satisfy the boundary conditions \eqref{eq 6} with $z_0=0$ are $Z_n(z)=B_n \cos(\lambda_n z)$ where $\lambda_n=\frac{n\pi}{h}$, and $B_n$ is a constant. Consequently, starting from the superposition principle, and from \eqref{eq 8}, we obtain
\begin{equation}
\label{aeq 15}
\overline{c^{y}}(x,z)=B_0+\sum_{n=1}^\infty B_n \cos(\lambda_n z) E_{\alpha}\left(-\kappa\lambda_n^2 x^\alpha\right).
\end{equation}
Finally, by imposing the boundary condition \eqref{eq 7}, and using the identity
\begin{equation}
\label{aeq q6}
\delta(x-a)=\frac{1}{d}+\frac{2}{d}\sum_{n=1}^\infty \cos\left(\frac{n\pi a}{d}\right)\cos\left(\frac{n\pi x}{d}\right),
\end{equation}
we obtain \eqref{eq 19}.

\section{Solution of the $\alpha$--SM model}

In order to solve the equations \eqref{eq 9} and \eqref{eq 10}, we use Frobenius method. Substituting equation \eqref{Aeq 12} in equation \eqref{eq 9} we obtain,
\begin{equation}\label{Aeq 13}
 \frac{b_{0}}{\Gamma(\beta)}{_0^C D^{\alpha}_x} x^{p} + \frac{b_{1}x^{p}}{\Gamma(\beta)} + \displaystyle\sum_{j=1}^{\infty}\Big[\frac{b_{j+1}}{\Gamma(j \alpha + \beta)} + \tau \lambda^{2} \frac{b_{j-1}}{\Gamma((j-1)\alpha + \beta)} \Big] x^{j \alpha + p} = 0.
\end{equation}
As in Appendix A, if $p \neq 0$ the Caputo fractional derivative for the function $X(x^{\alpha})$ is not defined. Therefore, $p = 0$ and $\beta = 1$. In this case, the equation \eqref{Aeq 13} can be written as,
\begin{equation}\label{Aeq 14}
 \frac{b_{1}}{\Gamma(1)} + \displaystyle\sum_{j=1}^{\infty}\Big[\frac{b_{j+1}}{\Gamma(j \alpha + 1)} + \tau \lambda^{2} \frac{b_{j-1}}{\Gamma((j-1)\alpha + 1)} \Big] x^{j \alpha} = 0.
\end{equation}
From the equation \eqref{Aeq 14} we have $b_{1} = 0$ and $b_{2j+1} = 0$. Furthermore, for $b_{2j}$ we get
\begin{equation}\label{Aeq 15}
b_{2j}=\frac{(-1)^{j}\displaystyle\prod_{i=1}^{j}\Gamma((2i-1)\alpha + 1)}{\displaystyle\prod_{i=1}^{j}\Gamma(2(i-1)\alpha+1)}(\tau \lambda^{2})^{j}b_{0}.
\end{equation}
By substituting \eqref{Aeq 15} in \eqref{Aeq 10} (for $b_{0} = 1)$ we get the equation \eqref{eq 11},
\begin{equation}\label{Aeq 16}
X(x^{\alpha})=\displaystyle\sum_{j=0}^{\infty} (-1)^{j}\frac{\displaystyle\prod_{i=1}^{j}\Gamma((2i-1)\alpha+1)}{\displaystyle\prod_{i=1}^{j+1}\Gamma(2(i-1)\alpha+1)}(\tau \lambda^{2}x^{2\alpha})^{j}.
\end{equation}

On the other hand, \eqref{eq 10} can also be solved by Frobenius method.
We consider the following power series as analytic solution of \eqref{eq 10}
\begin{equation}\label{Aeq 18}
\begin{array}{lll}
Z(z) &=& \displaystyle\sum^\infty_{n=0}c_n \Big(z^{1+\frac{k}{2}}\Big)^{n+r}. \\
\end{array}
\end{equation}
Substituting $Z(z)$ into \eqref{eq 10} we have (shifting the indices of first two series so all terms are of the form $z^{n+r}$),
\begin{equation}\label{Aeq 19}
\begin{array}{lll}
 \frac{1}{4}(2+k)r(-2+(2+k)r)c_0\Big(z^{1+\frac{k}{2}}\Big)^{r-2}+
 \frac{1}{4}(2+k)(1+r)(k+(2+k)r)c_1\Big(z^{1+\frac{k}{2}}\Big)^{r-1}+ \\
 \displaystyle\sum^\infty_{n=0}\Big[\frac{1}{4}(2+k)(2+n+r)(-2+(2+k)(n+2)+(2+k)r)c_{n+2}+\lambda^2 c_n \Big]\Big(z^{1+\frac{k}{2}}\Big)^{n+r}=0.
\end{array}
\end{equation}
All coefficients of powers $\Big(z^{1+\frac{k}{2}}\Big)^{n+r}$ must equate to zero to obtain a solution. This yields two cases. In the first case ($c_1 = 0$ and $r=\fracc{2}{2+k}$) the solution is given by,
\begin{equation}\label{Aeq 20}
\begin{array}{lll}
 Z_1(z) = \displaystyle\sum^\infty_{n=0}\frac{(-1)^n(\frac{1}{2})^{2n+\frac{1}{2+k}}(\frac{2\lambda}{2+k})^{2n+\frac{1}{2+k}}}{n!\Gamma(1+\frac{1}{2+k}+n)} (z^{1+\frac{k}{2}})^{2n+\frac{2}{2+k}} = z^{\frac{1}{2}}J_{\frac{1}{2+k}}\left(\frac{2\lambda}{2+k} z^{1+\frac{k}{2}}\right).
\end{array}
\end{equation}
In the second case ($c_0 = 0$ and $r=-1$), the solution is
\begin{equation}\label{Aeq 21}
\begin{array}{lll}
 Z_2(z) = \displaystyle\sum^\infty_{n=0}\frac{(-1)^n(\frac{1}{2})^{2n-\frac{1}{2+k}}(\frac{2\beta}{2+k})^{2n-\frac{1}{2+k}}}{n!\Gamma(1-\frac{1}{2+k}+n)} (z^{1+\frac{k}{2}})^{2n-1} = z^{\frac{1}{2}}J_{-\frac{1}{2+k}}\left(\frac{2\lambda}{2+k} z^{1+\frac{k}{2}}\right).
\end{array}
\end{equation}
Using the superposition principle we have that
\begin{equation}\label{Aeq 22}
\begin{array}{lll}
 Z(z) =  A Z_1(z) + B Z_2(z) =  Az^{\frac{1}{2}}J_{\frac{1}{2+k}}\left(\frac{2\lambda}{2+k} z^{1+\frac{k}{2}}\right) + Bz^{\frac{1}{2}}J_{-\frac{1}{2+k}}\left(\frac{2\lambda}{2+k} z^{1+\frac{k}{2}}\right).
\end{array}
\end{equation}
 where $A$ and $B$ are constants.
Applying the boundary condition  $Z'(0) = 0$ \eqref{eq 6} in \eqref{Aeq 22} we obtain $A=0$. On the other hand, by applying the boundary condition  $Z'(h) = 0$ \eqref{eq 6} in \eqref{Aeq 22} with $A=0$ we obtain
\begin{equation}\label{Aeq 25}
\begin{array}{lll}
 &  &h^{\frac{1}{2\mu}-1}\displaystyle\sum^\infty_{n=1}\fracc{(-1)^n(\frac{1}{2})^{2n-1-\mu}(2 \lambda \mu)^{2n-\mu}}{(n-1)!\Gamma(1-\mu+n)} (h^{\frac{1}{2\mu}})^{2n-1} =0.
\end{array}
\end{equation}
Making the change of variable $m=n-1$ we get
\begin{equation}\label{Aeq 26}
\begin{array}{lll}
 &  &\displaystyle\sum^\infty_{m=0}\fracc{(-1)^{m}(\frac{1}{2})^{2m+1-\mu}(2 \lambda \mu)^{2m+1-\mu}}{m!\Gamma(1-\mu+1+m)} (h^{\frac{1}{2\mu}})^{2m+1-\mu} =0.
\end{array}
\end{equation}
Therefore we should have
\begin{equation}\label{Aeq 27}
\begin{array}{lll}
 J_{1-\mu}\Big(2\lambda\mu h^{\frac{1}{2\mu}}\Big)&=&0.
\end{array}
\end{equation}
Consequently, $\lambda=\lambda_n$ is given by the zeros of the Bessel function of order $-\mu + 1$.

Finally, from the superposition principle, and from \eqref{eq 8}, \eqref{Aeq 16} and \eqref{Aeq 22}, the solution of \eqref{eq 5} can be written as
\begin{equation}\label{eqa 15}
C(x,z)=B_{0}+z^{\frac{1}{2}}\sum_{n=1}^{\infty}B_{n}J_{-\mu}(2 \lambda_{n} \mu z^{\frac{1}{2\mu}})\sum_{j=1}^{\infty} (-1)^{j}\frac{\prod_{i=1}^{j}\Gamma[(2i-1)\alpha+1]}{\prod_{i=1}^{j+1}\Gamma[2(i-1)\alpha+1]}(\tau \lambda_{n}^{2}x^{2\alpha})^{j}.
\end{equation}
In order to fix the constants $B_n$, we use the boundary condition  $C(0,z)=\frac{Q}{\gamma z^k}\delta(z-H_{s})$ \eqref{eq 7} in \eqref{eqa 15}. We consider
\begin{equation}\label{Aeq 29}
 B_{0}+z^{\frac{1}{2}}\sum_{n=1}^{m}B_{n}J_{-\mu}(2 \lambda_{n} \mu z^{\frac{1}{2\mu}}) =\frac{Q}{\gamma z^k}\delta(z-H_{s})
\end{equation}
as a good approximation because $J_{-\mu}(2 \lambda_{n} \mu z^{\frac{1}{2\mu}})$ goes to zero when $n\rightarrow \infty$.
Now, making
\begin{equation}\label{Aeq 30}
\int^h_0 \big(B_{0}+z^{\frac{1}{2}}\sum_{n=1}^{m}B_{n}J_{-\mu}(2 \lambda_{n} \mu z^{\frac{1}{2\mu}})\Big)J_{-\mu}(2 \lambda_{l} \mu z^{\frac{1}{2\mu}}) dz=\int^h_0\frac{Q}{\gamma z^k}\delta(z-H_{s})J_{-\mu}(2 \lambda_{l} \mu z^{\frac{1}{2\mu}})dz,
\end{equation}
where $l\in\{0,1,...,m\}$, we fix $B_n$ from the numerical solution of the following system
\begin{equation}\label{Aeq 31}
 B_{0}\int^h_0 J_{-\mu}(2 \lambda_{l} \mu z^{\frac{1}{2\mu}}) dz+\sum_{n=1}^{m}B_{n}\int^h_0 z^{\frac{1}{2}}J_{-\mu}(2 \lambda_{n} \mu z^{\frac{1}{2\mu}})J_{-\mu}(2 \lambda_{l} \mu z^{\frac{1}{2\mu}}) dz=\frac{Q}{\gamma H_s^k}J_{-\mu}(2 \lambda_{l} \mu H_s^{\frac{1}{2\mu}}).
\end{equation}


\end{document}